\documentclass[%
 reprint,%
 amssymb, amsmath,%
 aip,cha,%
]{revtex4-1}

\usepackage{graphics}
\usepackage{amsmath}
\usepackage{amsfonts}
\usepackage{latexsym}
\usepackage{amsbsy}
\usepackage{epstopdf}
\usepackage{bm}%

\newcommand{\ket}[1]{\vert #1 \rangle}

\usepackage{color}

\usepackage{ulem} 

\newcommand{\dyadic}[1]{{#1}
\setbox0=\hbox{$\scriptstyle\leftrightarrow$}
   \setbox2=\hbox{$#1$}
   \dimen0=.5\wd0 \advance\dimen0 by-.5\wd2
   \advance\dimen0 by-\wd0
   \kern\dimen0
{^{\hbox{$\scriptstyle\leftrightarrow$}}}}

\begin{document}

\title{A New Quantum-Based Power Standard: Using Rydberg Atoms for a SI-Traceable Radio-Frequency Power Measurement Technique in Rectangular Waveguides}
\thanks{Publication of the U.S. government, not subject to U.S. copyright.}
\author{Christopher~L.~Holloway}
\email{christopher.holloway@nist.gov}
\author{Matthew T. Simons}
\author{Marcus D. Kautz}
\author{Abdulaziz H. Haddab}
\author{Joshua A. Gordon}
\affiliation{National Institute of Standards and Technology (NIST), Boulder,~CO~80305}
\author{Thomas P. Crowley}
\affiliation{Xantho Technologies, LLC in Madison, WI, 53705}

\date{\today}

\begin{abstract}
In this work we demonstrate an approach for the measurement of radio-frequency (RF) power using electromagnetically induced transparency (EIT) in a Rydberg atomic vapor.  This is accomplished by placing alkali atomic vapor in a rectangular waveguide and measuring the electric (E) field strength (utilizing EIT and Autler-Townes splitting) for a wave propagating down the waveguide. The RF power carried by the wave is then related to this measured E-field, which leads to a new direct International System of Units (SI) measurement of RF power. To demonstrate this approach, we first measure the field distribution of the fundamental mode in the waveguide and then measure the power carried by the wave at both 19.629 GHz and 26.526 GHz. We obtain good agreement between the power measurements obtained with this new technique and those obtained with a conventional power meter.
\end{abstract}

\maketitle

The world of measurement science is changing rapidly due to the International System of Units (SI) redefinition planned for late 2018. As a result of the shift towards fundamental physical constants, the role of primary standards must change. This includes radio-frequency (RF) power. The current method of power traceability is typically based on an indirect path through a thermal measurement using a calorimeter, in which temperature rise created by absorbed microwave energy is compared to the DC electrical power. A direct SI-traceable measurement of RF power is desired and to accomplish this we will utilize recent work on electric (E) field metrology using Rydberg atomic vapor.

\begin{figure}[!t]
\scalebox{.080}{\includegraphics*{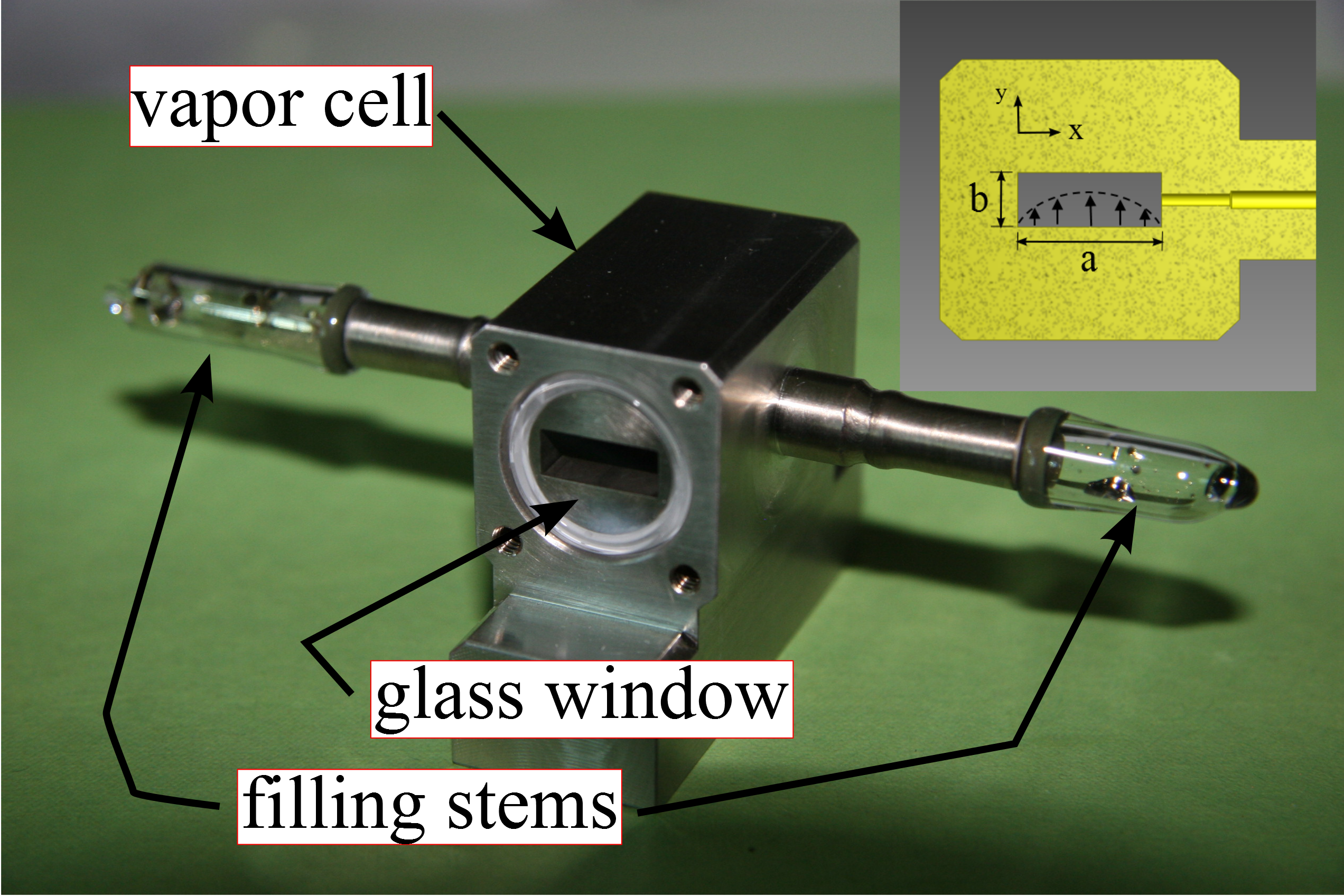}}
\caption{WR-42 rectangular waveguide vapor cell with waveguide dimensions. The vapor cell constist of a 34-mm section of waveguide with glass windows attached to each end (and filled with $^{133}$Cs.)}
\label{fig1}
\end{figure}

It can be shown that the E-field of the fundamental mode (the transverse electric ($TE_{10}$) mode) in the rectangular waveguide, shown in Fig.~\ref{fig1}, is given by \cite{johnk}
\begin{equation}
{\bf{E}}=E_0\sin\left(\frac{\pi}{a}\,x\right) {\bf{a}}_y
\label{e1}
\end{equation}
and the power carried by this mode is
\begin{equation}
P = E^2_0\,\,\frac{ab}{4}\,\sqrt{\frac{\epsilon_0}{\mu_0}}\,\sqrt{1-\left(\frac{c}{2\,a\,f}\right)^2} \,\,\, , \label{p1}
\end{equation}
where $E_0$ is the amplitude of the E-field at the center of the waveguide, $a$ and $b$ are the dimensions of the waveguide, $f$ is the frequency, $\epsilon_0$ and $\mu_0$ are the permittivity and permeability of free space, and $c$ is the speed of light {\it in vacuo}.

If $E_0$ can be determined, then the power can be measured. We can leverage the recent works in the development of a new atom-based, SI-traceable, approach for determining E-field strengths, in which significant progress has been made in the development of a novel Rydberg-atom spectroscopic approach for RF E-field strength measurements \cite{r1, r2, r3, r5, r6a, r7, r8, r8b, fan2, dave1, dave2}.  This approach utilizes the phenomena of electromagnetically induced transparency (EIT) and Autler-Townes (AT) splitting \cite{r6a, r1,r2, EIT_Adams}, and can lead to a direct SI traceable, self-calibrated measurement.

\begin{figure}[!t]
\centering
\scalebox{.30}{\includegraphics*{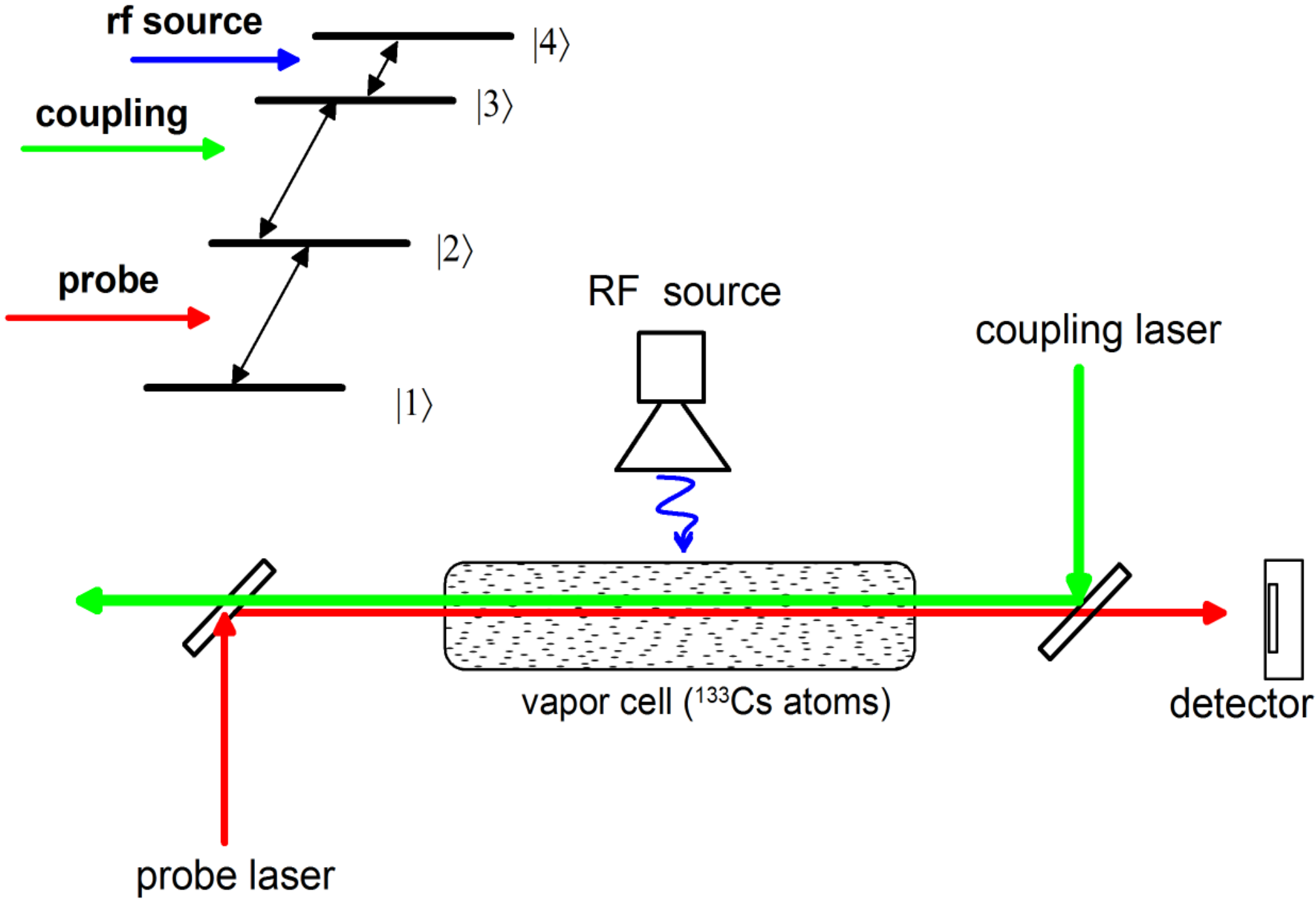}}
\caption{Illustration of a four-level system, and the vapor cell setup for measuring EIT, with counter-propagating probe and coupling beams.}
\label{4level}
\end{figure}

Consider a sample of stationary four-level atoms illuminated by a single weak (``probe") light field, as depicted in Fig~\ref{4level}. In this approach, one laser is used to probe the response of the atoms and a second laser is used to excite the atoms to a Rydberg state (the “coupling” laser). In the presence of the coupling laser, the atoms become transparent to the probe laser transmission (this is the concept of EIT). The coupling laser wavelength is chosen such that the atom is at a high enough state such that an RF field can cause an atomic transition of the atom. The RF transition in this four-level atomic system causes a splitting of the transmission spectrum (the EIT signal) for a probe laser.  This splitting of the probe laser spectrum is easily measured and is directly proportional to the applied RF E-field amplitude (through Planck's constant and the dipole moment of the atom). By measuring this splitting ($\Delta f_m$), we get a direct measurement of the magnitude of the RF E-field strength for a time-harmonic field from \cite{r1}:
\begin{equation}
	|E| = 2 \pi \frac{\hbar}{\wp}\Delta f_m \quad ,
	\label{mage2}
\end{equation}
where $\hbar$ is Planck's constant, $\wp$ is the atomic dipole moment of the RF transition (see Ref. \cite{r1, dual} for discussion on determining $\wp$ and values for various atomic states), and $\Delta f_m$ is the measured splitting when the coupling laser is scanned.  If the probe laser is scanned a Doppler mismatch correction is needed in this expression \cite{EIT_Adams, linear}. We consider this type of measurement of the E-field strength a direct, SI-traceable, self-calibrated measurement in that it is related to Planck's constant (which will become a SI quantity defined by standard bodies in the near future) and only requires a frequency measurement ($\Delta f_m$, which can be measured very accurately and is calibrated to the hyperfine atomic structure).

\begin{figure}
\centering
\scalebox{.26}{\includegraphics*{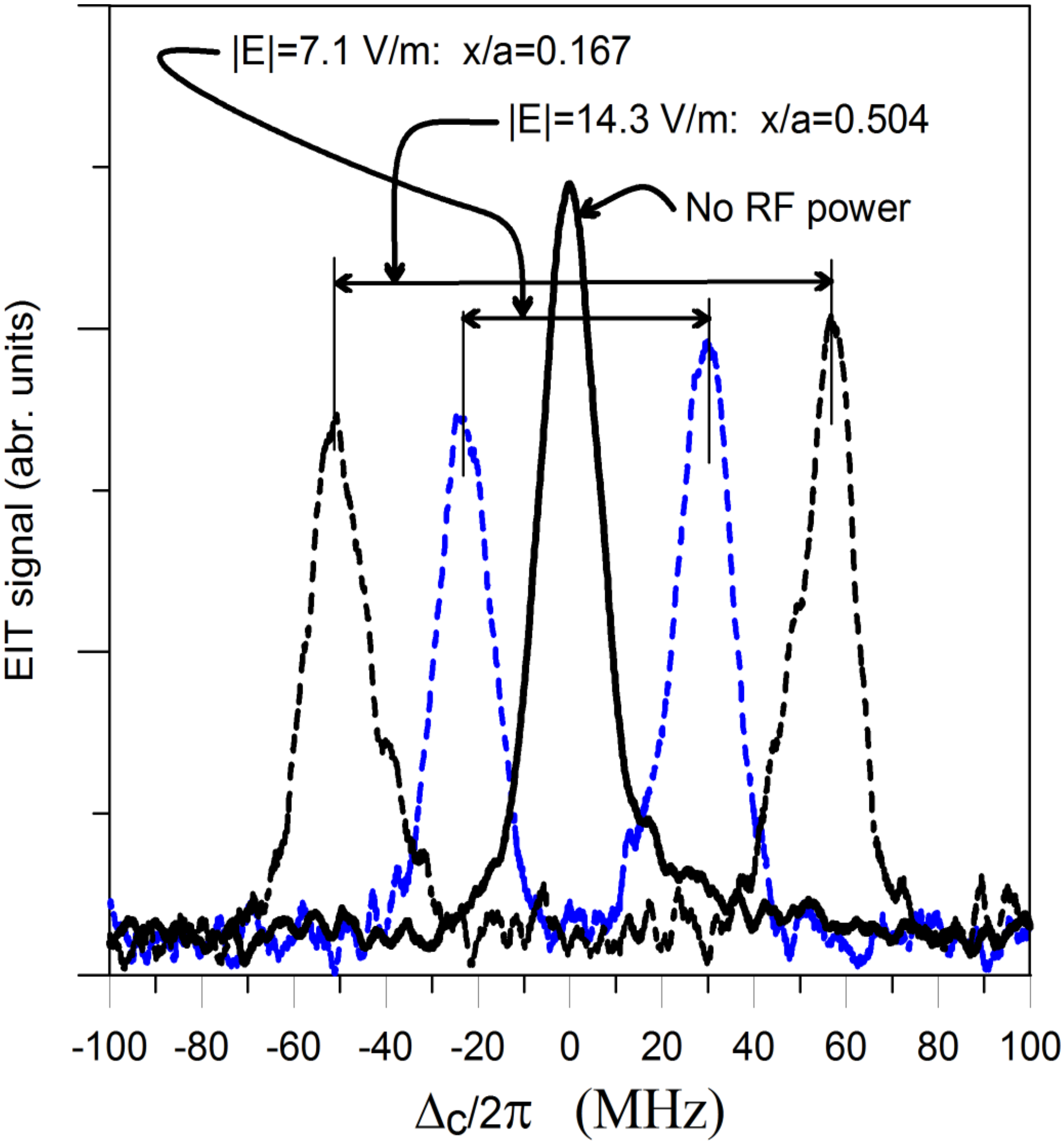}}
\caption{Illustration of the EIT signal (i.e., probe laser transmission through the cell) as a function of coupling laser detuning $\Delta_c$. This dataset is for 19.629~GHz and corresponds to this following 4-level atomic system: $6_{1/2}-6P_{3/2}-34D_{5/2}-35P_{3/2}$. The dashed curves correspond to two different $x$-locations across the WR42 waveguide for an input power of -20.76~dBm.}
\label{EIT}
\end{figure}

A typical measured spectrum for an RF source with different E-field strength is shown in Fig.~\ref{EIT}. This figure shows the measured EIT signal for two E-field strengths (more details on these results are given below). In this figure, $\Delta_c$ is the detuning of the coupling laser  (where $\Delta_c=\omega_c-\omega_o$; $\omega_o$ is the on-resonance angular frequency of the Rydberg state transition and $\omega_c$ is the angular frequency of the coupling laser). Notice that the AT splitting increases with increasing applied E-field strength. To obtain these results, we use cesium atoms ($^{133}$Cs) and the levels $\ket{1}$, $\ket{2}$, $\ket{3}$, and $\ket{4}$ correspond respectively to the $^{133}$Cs  $6S_{1/2}$ ground state,  $6P_{3/2}$ excited state, and two Rydberg states.  The probe is locked to the D2 transition (a $852$~nm laser).  The probe beam is focused to a full-width at half maximum (FWHM) of 290~$\mu$m, with a power of 3.2~$\mu$W. To produce an EIT signal, we apply a counter-propagating coupling laser (wavelength $\lambda_c \approx 510$~nm) with a power of 17.3~mW, focused to a FWHM of 380~$\mu$m.  The coupling laser was scanned across the  $6P_{3/2}$ -- $34D_{5/2}$ Rydberg transition ($\lambda_c=511.1480$~nm). We modulate the coupling laser amplitude with a 30~kHz square wave and detect any resulting modulation of the probe transmission with a lock-in amplifier. This removes the Doppler background and isolates the EIT signal, as shown in the solid curve of Fig.~\ref{EIT}. Application of RF (details below) at 19.629~GHz to couple states $34D_{5/2}$ and $35P_{3/2}$ splits the EIT peak as shown in the dashed curves in the figure.  We measure the frequency splitting of the EIT peaks in the probe spectrum $\Delta f_m$ and determine the E-field amplitude using (\ref{mage2}) as shown in Fig.~\ref{EIT}. For these measurement, the dipole moment for the resonant RF transition is $\wp=723.393 e a_0$ (which includes a radial part of $1476.619 e a_0$ and an angular part of $0.48989$, which correspond to co-linear polarized optical and RF fields, where $e$ is the elementary charge; $a_0=0.529177\times 10^{-10}$~m and is the Bohr radius).

\begin{figure*}[!t]
\centering
\scalebox{.45}{\includegraphics*{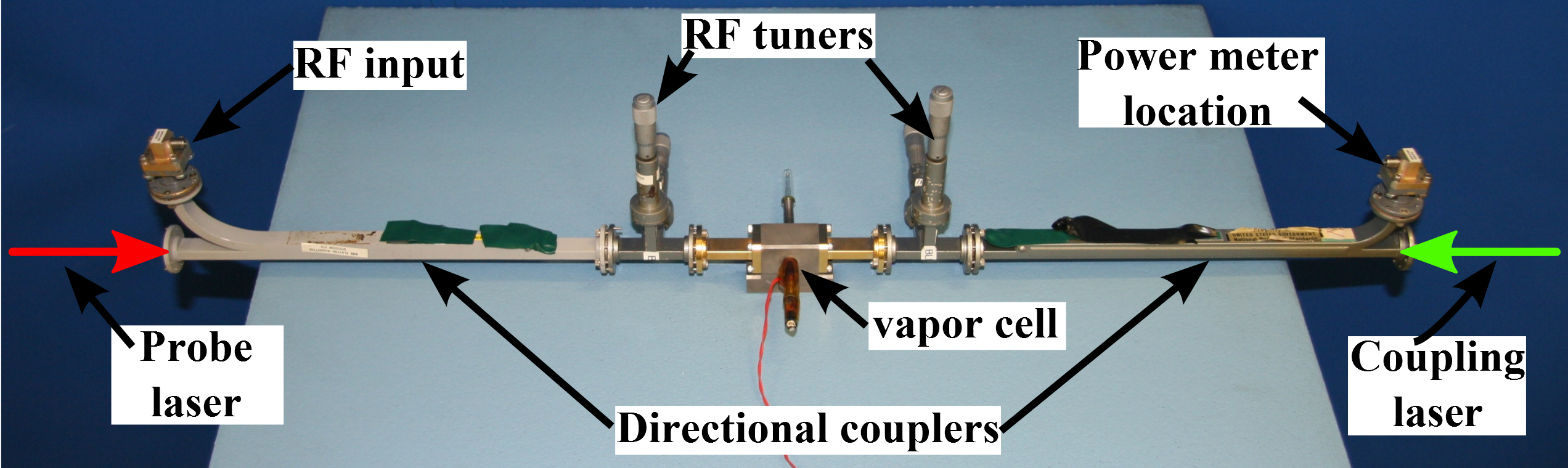}}
\caption{Photos of experimental setup for vapor-cell filled waveguide.}
\label{setup}
\end{figure*}

In order to measure the the power propagating down a WR42 rectangular waveguide, we placed a $^{133}$Cs vapor cell in the waveguide system shown in Fig.~\ref{setup}. The experimental setup includes two directional couplers, two RF tuners, and a 34~mm section of waveguide that serves as the vapor cell. The vapor cell consists of a 34-mm length of WR42 stainless-steel waveguide with glass windows attached to each end (attached with vacuum epoxy), see Figs.~\ref{fig1} and \ref{setup}.  The glass windows allow the vapor-cell waveguide to be filled with $^{133}$Cs under vacuum. The directional couplers were used to allow the probe and coupling laser to propagate down the waveguide system and interact with the $^{133}$Cs vapor, while at the same time allowing RF power to be coupled into the waveguide system (the directional coupler on the left) and allowing RF power to be coupled out of the waveguide system (the directional coupler on the right). The output of this second directional coupler was attached to a conventional RF power meter.  The presence of the two windows on the vapor cell results in the possibility of RF standing waves inside the vapor-cell along the propagation direction (along the waveguide axis).  The RF tuners are used to minimize and eliminate these standing waves (discussed below).

\begin{figure}
\centering
\scalebox{.26}{\includegraphics*{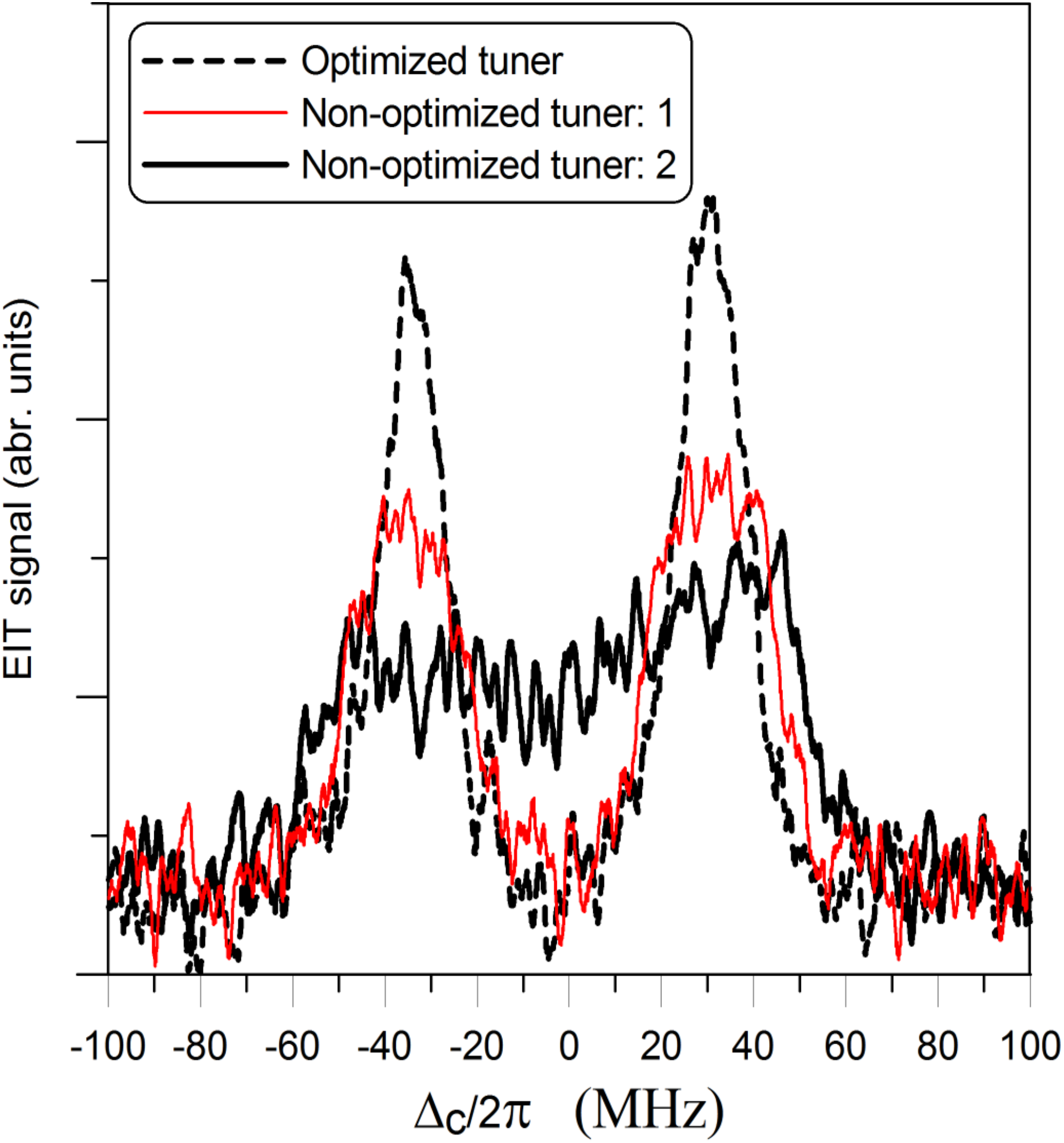}}
\caption{The effects of the standing waves (inhomogeneous field) on the EIT line width. These results are for $x/a=0.5$, 19.629~GHz, and an input power of -24.79~dBm.}
\label{linew}
\end{figure}

The WR42 waveguide system has dimensions of $a=10.668$~mm and $b=4.318$~mm which allows for only one propagating mode (the fundamental mode) between 18~GHz and 27~GHz. Thus, we perform experiments for two frequencies in this range, i.e., 19.629~GHz and 26.526~GHz.  We first perform experiments at 19.629~GHz which correspond to the $6S_{1/2}-6P_{3/2}-34D_{5/2}-35P_{3/2}$ atomic system. The waveguiding system was placed on a translation-stage, which allowed the probe and coupling lasers to be scanned (while maintaining their counter-propagation alignment) across the $x$-axis of the waveguide. The EIT signal at two different $x$-axis locations in the waveguide is shown in Fig.~\ref{EIT}. These results are for an input power (input to the directional coupler on the left, see Fig.~\ref{setup}) of -20.76~dBm.  As discussed above, the presence of the glass windows can result in possible standing waves inside the vapor-cell. In order to get an accurate measurement for the forward propagating power, these standing waves needed to be eliminated (or at least minimized as much as possible).  We can use the linewidth of the EIT signal as a means of determining when the standing waves (SWs) effect is minimized.  The SWs can result in a broadening of the EIT linewidth, a direct result of the inhomogeneous E-field variation (due to the SWs) along the propagation direction \cite{dave1}. An inhomogeneous E-field along the direction of the laser beam propagation can cause a broadening of the EIT linewidth.  To minimize this effect, we varied the RF tuners on both sides of the vapor-cell waveguide until the EIT linewidth was minimized, which was an indication when the RF SWs in the vapor-cell were minimized.  The effect of the SWs on the EIT linewidth is shown in Fig.~\ref{linew}, where we show three EIT signals. One of the EIT signals is for the case when the RF tuners are optimized and the other two EIT signals are for the case when the RF tuners are non-optimized.  We see that the EIT linewidth for the non-optimized cases is larger than the optimized case. Furthermore, the EIT signal shown in Fig.~\ref{EIT} is for the optimized tuners and we see that for this optimized case, the EIT linewidth is approximately the same as the case with no RF power in the waveguide, indicating that the RF SWs in the waveguide are minimized.

\begin{figure}
\centering
\scalebox{.27}{\includegraphics*{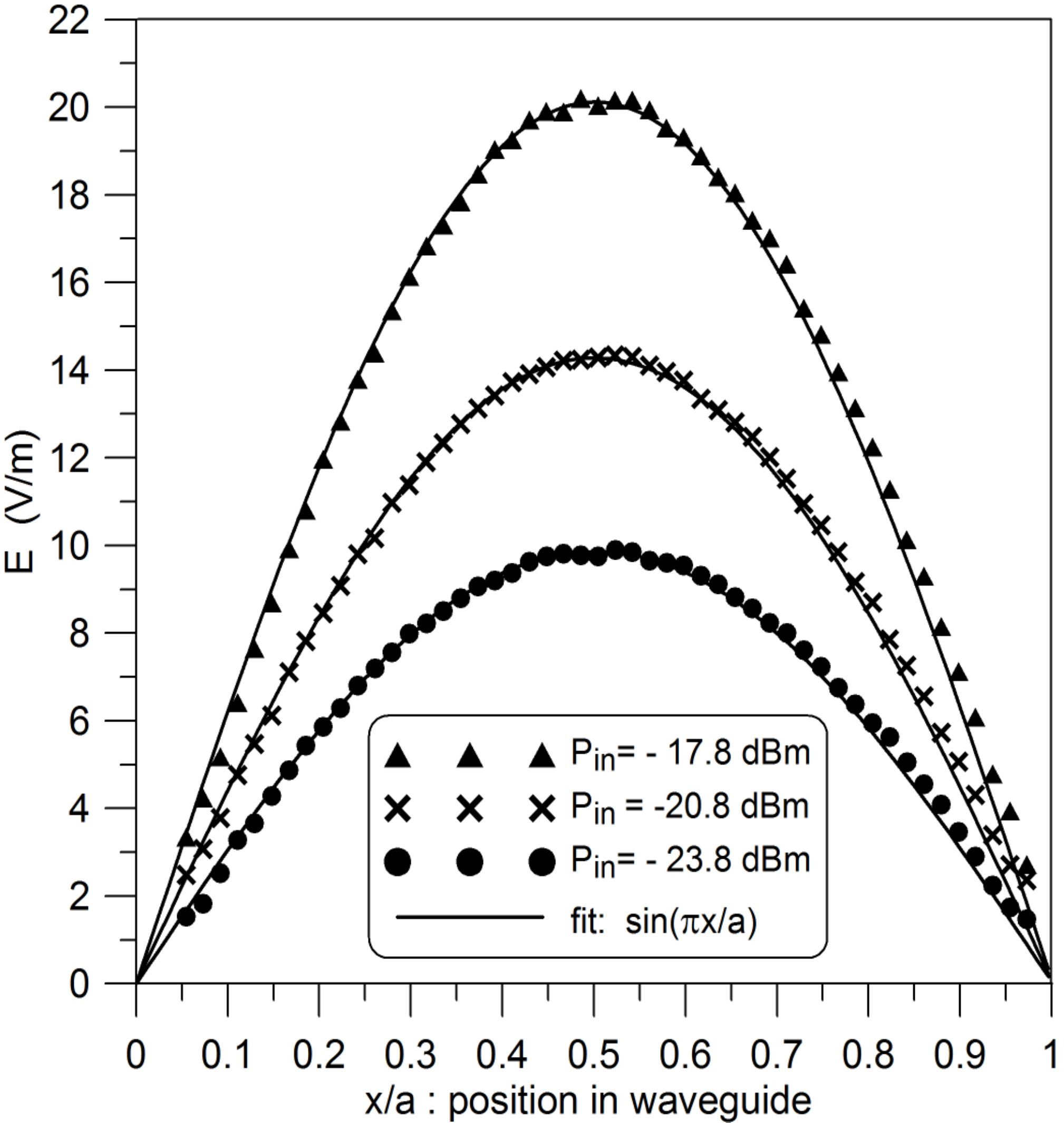}}
\caption{E-field distribution long the $x$-axis of the waveguide at 19.629~GHz.}
\label{scan}
\end{figure}

We next measure the E-field distribution across the $x$-axis in the waveguide for different input RF power levels. This is done by scanning the laser across the $x$-axis of the waveguide from $x=0$ to $x=a$ (actually scanning the waveguide system via the translation stage).  The measured E-field distributions inside the waveguide for three different input powers (input to the directional coupler) are shown in Fig.~\ref{scan}.  To obtain the results, we first measured $\Delta f_m$ of the EIT signal at different $x$ locations, then using eq.~(\ref{mage2}), the E-field strength was determined. As indicated from eq.~(\ref{e1}), the E-field dependance should follow a $\sin(\pi x/a)$ distribution for the $TE_{10}$ mode. The results in this figure indicate that the measured E-field distribution inside the waveguide follows this behavior very well.

\begin{figure}[!h]
\centering
\scalebox{.28}{\includegraphics*{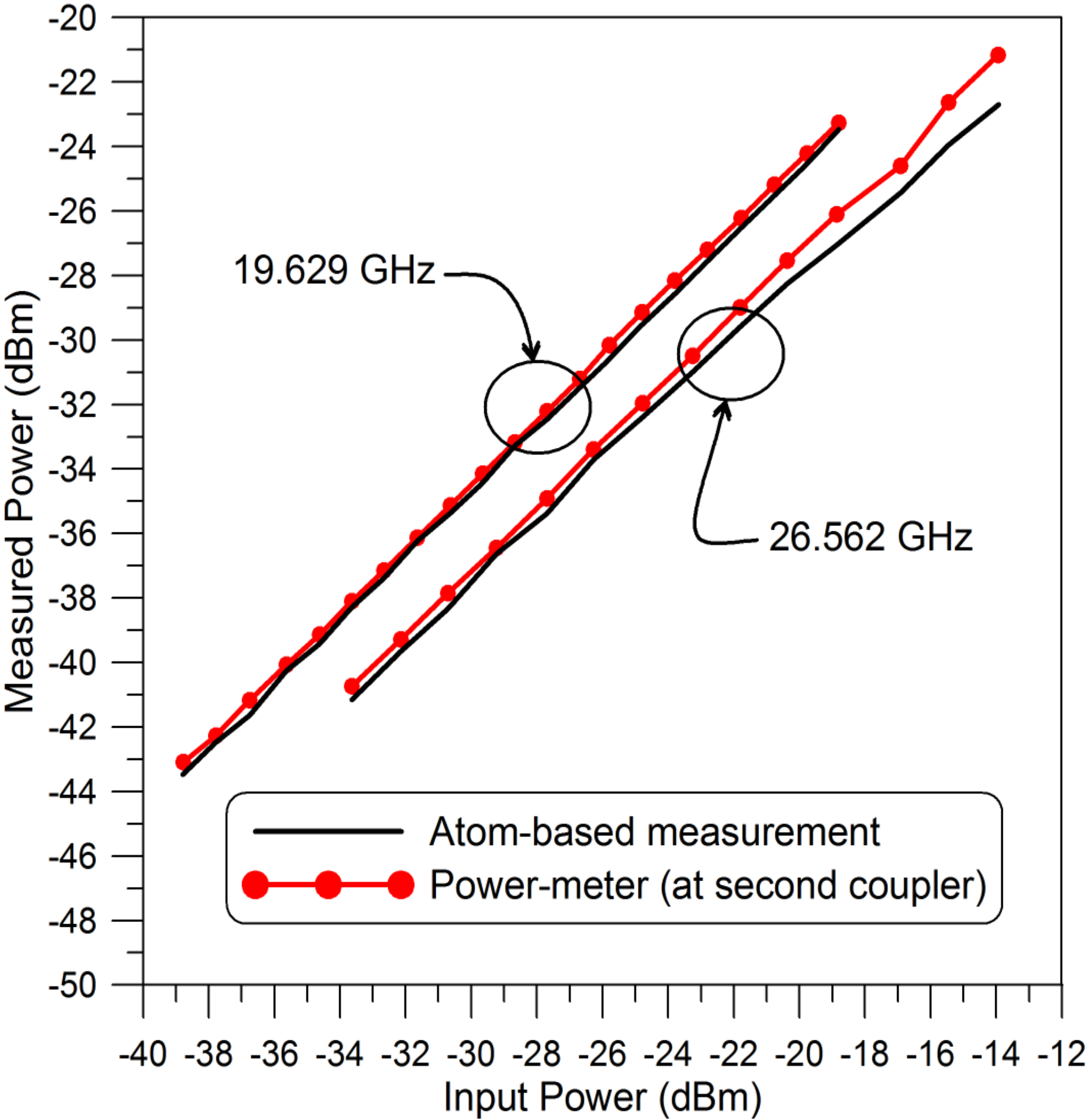}}
\caption{Measurements power in the waveguide versus input power at both 19.629~GHz and 26.526~GHz.}
\label{power}
\end{figure}

With the E-field strength determined at the center of the waveguide (i.e., $x=a/2$), eq.~(\ref{p1}) can be used to determine the power flowing down the waveguide system.  Fig.~\ref{power} shows the measured RF power in the waveguide as a function of input power (i.e., the input power at the directional coupler on the left). These results are at 19.629~GHz and for a 4-level atomic system ($6S_{1/2}-6P_{3/2}-34D_{5/2}-35P_{3/2}$) and with the same probe and coupling laser bandwidth and powers as that used above.  As a comparison we also show results obtained from a conventional power meter connected to the right directional coupler.  The power-meter results were corrected for the losses in the waveguide system (i.e., loss and directional coupler attenuation). The comparison shows very good agreement.

We performed a second set of measurements at 26.526~GHz. These experiments correspond to the following 4-level atomic system: $6S_{1/2}-6P_{3/2}-31D_{5/2}-32P_{3/2}$.  Once again the probe laser was locked to the D2 $^{133}$Cs transient (a $852$~nm laser) and the coupling laser was scanned across the  $6P_{3/2}$ -- $31D_{5/2}$ Rydberg transition ($\lambda_c=511.787$~nm). The power and beamwidth for probe and coupling were the same as used for 19.629~GHz.  We first measured the E-field along the $x$-axis for the waveguide. While the results are not shown here, the results are similar to those for the 19.629~GHz case, i.e., following the expected $\sin(\pi x/a)$ behavior.  With the E-field strength determined [using $\wp=592.158 e a_0$ (which includes a radial part of $1208.737 e a_0$ and an angular part of $0.48989$)]  in the center of the waveguide (i.e., $x=a/2$), eq.~(\ref{p1}) can be used to determine the power flowing down the waveguide system.  Fig.~\ref{power} shows the measured RF power in the waveguide as a function of input power (i.e., the input power at the directional coupler on the left). Also, shown are the results from a conventional power-meter, and good agreement is shown once again.

In the search for a new quantum-based power standard, we have presented a fundamental new SI-traceable method for measuring RF power.  The technique is based on Rydberg atomic vapor placed in rectangular waveguide and utilizing the EIT/AT approach.  We first demonstrated the ability to measure the E-field distribution of the fundamental $TE_{10}$ mode in the waveguide.  We then performed measurements of RF power from the Rydberg-atom approach and compared it to results obtained from a conventional power meter, and demonstrated very good agreement.
The results for both the 19.629~GHz and 26.526~GHz cases demonstrate the ability of this new approach to measure RF power inside a waveguide, and can lead to a new direct SI-traceable approach for power metrology.  While the uncertainty of this new measurement technique is currently being investigated, when compared to conventional power metrology approaches, this new approach: (1) is a more direct SI traceable approach, (2) has the possibility of having much lower uncertainty, (3) exhibits much better frequency range, and (4) has much better dynamic range (i.e., power-level ranges).

\end{document}